# One-shot Generative Prior in Hankel-k-space for Parallel Imaging Reconstruction

Hong Peng, Chen Jiang, Jing Cheng, Minghui Zhang, Shanshan Wang, *Senior Member, IEEE*,
Dong Liang, *Senior Member, IEEE,* Qiegen Liu, *Senior Member, IEEE*

*Abstract*—Magnetic resonance imaging serves as an essential tool for clinical diagnosis. However, it suffers from a long acquisition time. The utilization of deep learning, especially the deep generative models, offers aggressive acceleration and better reconstruction in magnetic resonance imaging. Nevertheless, learning the data distribution as prior knowledge and reconstructing the image from limited data remains challenging. In this work, we propose a novel Hankel-k-space generative model (HKGM), which can generate samples from a training set of as little as one k-space data. At the prior learning stage, we first construct a large Hankel matrix from k-space data, then extract multiple structured k-space patches from the Hankel matrix to capture the internal distribution among different patches. Extracting patches from a Hankel matrix enables the generative model to be learned from the redundant and low-rank data space. At the iterative reconstruction stage, the desired solution obeys the learned prior knowledge. The intermediate reconstruction solution is updated by taking it as the input of the generative model. The updated result is then alternatively operated by imposing low-rank penalty on its Hankel matrix and data consistency constraint on the measurement data. Experimental results confirmed that the internal statistics of patches within a single k-space data carry enough information for learning a powerful generative model and provide state-of-the-art reconstruction.

## I. Introduction

Magnetic resonance imaging (MRI) is a widely used non-invasive imaging technique for clinical diagnosis and research because of the excellent spatial resolution and soft-tissue contrast illustration. However, the acquisition speed is fundamentally limited due to hardware and physiological constraints and the requirement to satisfy Nyquist sampling rate. Long acquisition time is a burden for patients and makes MRI susceptible to motion artifacts. In this study, we focus on image reconstruction for accelerating parallel MRI (pMRI) [1].

This work was supported in part by National Natural Science Foundation of China under 61871206, 62161026. (H. Peng and C. Jiang are co-first authors.) (Corresponding authors: D. Liang and Q. Liu)

H. Peng, C. Jiang, M. Zhang and Q. Liu are with School of Information Engineering, Nanchang University, Nanchang 330031, China. ({penghong, jiangchen}@email.ncu.edu.cn, {zhangminghui, liuqiegen}@ncu.edu.cn)

J. Cheng, S. Wang, and D. Liang and are with Paul C. Lauterbur Research Center for Biomedical Imaging, SIAT, Chinese Academy of Sciences, Shenzhen 518055, China. (sophiasswang@hotmail.com, {dong.liang, jing.cheng}@siat.ac.cn)

In parallel imaging, multiple receive coils provide additional sensitivity information for successful reconstruction from incomplete sampling. The commonly used pMRI methods include generalized autocalibrating partial parallel acquisition (GRAPPA) [2] and sensitivity encoding (SENSE) [3]. GRAPPA works in k-space as an interpolation procedure, and SENSE works in image space using explicitly calculated coil sensitivity maps. The other successful approach for accelerating MRI is compressed sensing (CS) [4], which uses the sparsity prior and incoherent sampling. Later, the concept of low-rank modeling has shown increased popularity, such as the eigenvector-based iterative self-consistent parallel imaging reconstruction (ESPIRiT) [5], simultaneous autocalibrating and k-space estimation (SAKE) [6] (see Fig. 1(a)), low-rank matrix modeling of local k-space neighborhoods (LORAKS) [7] and annihilating filter-based low-rank Hankel matrix (ALOHA) [8]. Admittedly, the reconstruction accuracy of above-mentioned methods has large room to improve.

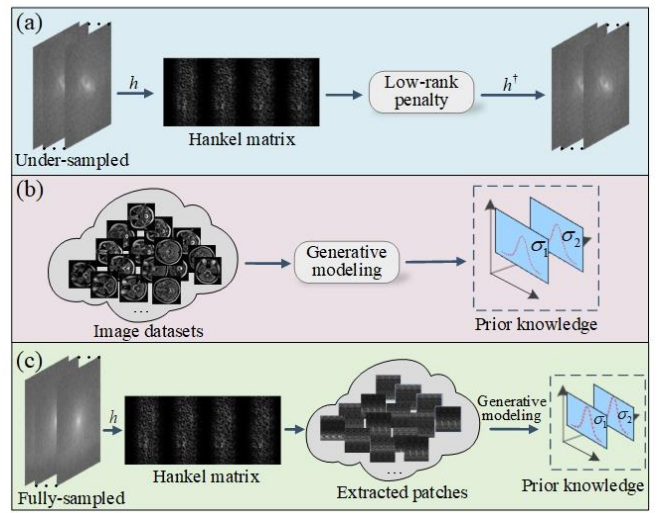

**Fig. 1.** Different prior knowledge methods for pMRI. (a) Conventional k-space iterative methods (e.g., SAKE) adopt low-rank penalty on Hankel matrix. (b) Existing generative modeling (e.g., HGGDP) on fully-sampled data. (c) The proposed generative modeling is conducted on a dataset of low-rank patches extracted from a single k-space measurement.

In comparison, deep learning (DL) methods are gaining popularity for their accuracy and efficiency. DL methods can be roughly categorized into the supervised manner and unsupervised manner. Supervised learning approaches [9]-[15] usually require thousands of fully-sampled data to train the network, which is difficult in some special circumstances. Unsupervised DL methodologies [16-21], such as deep generative models shown in Fig. 1(b), learn the distribution of objects and show great promise in accelerating MRI. For instance, Liu *et al.* [17] leveraged denoising autoencoder (DAE) as an explicit prior to address the highly under-sampled MRI reconstruction problem. Tezcan *et al.* [18] proposed an approach that learned the distribution of fully sampled magnetic resonance (MR) images and used it as an explicit prior term in reconstruction. A similar approach, PixelCNN [19], exploited a generative network as the image prior for MRI reconstruction. The generative flow (Glow) [20] that formulated in the latent space of invertible neural network-based generative models was proposed for reconstructing images from under-sampled MR data. Quan *et al.* [21] presented HGGDP for MRI reconstruction by utilizing the denoising score matching. They all used a certain amount of training data to learn the prior information for under-sampled MRI reconstruction.

Despite the promising progress of these deep generative models in MRI reconstruction, challenges remain. For example, training HGGDP for MRI reconstruction relies on 500 pieces of training data. Nevertheless, dataset collection is difficult and expensive, and training on large datasets often takes a long time. To alleviate the issues, we present a Hankel-k-space generative model (HKGM) that can learn to generate samples from a training set of as little as only one case of k-space data. Since directly training with one single sample is intractable, we conduct the generative model on a dataset of low-rank patches (see Fig. 1(c)). Specifically, we first construct a huge Hankel matrix from a k-space data, then extract multiple and low-rank Hankel structured k-space patches as the training dataset. Our proposal is motivated from the success of modelling the internal distribution of patches within a single natural image [22]-[30]. Classical examples include denoising [22]-[24], deblurring [25], super-resolution [26], outlier detection [27], and image editing [28].

The main contributions of this work are summarized as follows:
- To alleviate the issue of insufficient data samples in prior learning, we first construct a Hankel matrix from the sampled data in k-space, then extract multiple k-space patches from the matrix. Therefore, a large number of training data can be obtained within one case of k-space data.
- After the prior knowledge is learned, it is incorporated into the conditional generation for high-quality reconstruction. In addition to sample generation, we impose low-rank penalty on the Hankel matrix and data consistency constraint on the measurement data alternatively at each iteration. Since the learned prior has no restriction on the channel number, it can be adopted to pMRI reconstructions on different coils.

The rest of this paper is presented as follows. Section II briefly describes some reconstruction works using low-rank methodology and the research of score-based generative models. In Section III, we elaborate the theory of HKGM, including training and iteration processes. Section IV demonstrates experimental results and the comparison with state-of-the-arts. Section V concludes the present method.

## II. RELATED WORK

### A. Background on PMRI

The forward model for the multi-coil pMRI observation acquisition in k-space domain can be described as follows:
$$y_c = Mk_c + n_c, \ c = 1, 2, \cdots, C \quad (1)$$
where $M$ is a diagonal matrix whose diagonal elements are either 1 or 0 depending on the sampling pattern. $n_c$ is the noise. $k_c$ represents the $c$-th coil k-space data and $y_c$ denotes the corresponding acquired k-space measurement data. $C$ denotes the number of coils. The pMRI recovery can be formulated as an optimization problem:
$$\underset{k}{Min} \|Mk - y\|_2^2 + \lambda R(k) \quad (2)$$
where $\|Mk - y\|_2^2 = \sum_{c=1}^{C} \|Mk_c - y_c\|_2^2$ is the data fidelity term, which enforces data consistency with acquired k-space measurement $y$. $R(k)$ is the regularization term of $k$, $\|\cdot\|_2^2$ represents the $l_2$-norm and $\lambda$ is the factor that balances the data-fidelity term and the regularization term.

### B. Reconstruction using Low-rank Methodology

The structured low-rank matrix prior has shown promising results. For instance, a calibrationless parallel imaging reconstruction method named SAKE was presented by Peter *et al.* [6], which formulated the MRI reconstruction as a structured low-rank matrix completion problem. Empowered by recent k-space interpolation methods, Jin *et al.* [8] presented an annihilating filter-based low-rank Hankel matrix approach (ALOHA) as a general framework for sparsity-driven k-space interpolation, which combined pMRI with CS-MRI. Inspired by the recent mathematical discovery of a data-driven framelet that linked convolutional neural networks to Hankel matrix decomposition, Han *et al.* [31] demonstrated a fully data-driven DL algorithm for k-space interpolation. Subsequently, Zhang *et al.* [32] devised an approach named STDLR-SPIRiT, which combined the simultaneous two-directional low-rankness (STDLR) with the iterative self-consistent parallel imaging reconstruction (SPIRiT) to mine the data correlation from multiple receiver coils. Meanwhile, Haldar *et al.* [7] developed a flexible framework for constrained image reconstruction that used low-rank matrix modeling of local k-space neighborhoods (LORAKS). Soon later, the author introduced P-LORAKS [33], which extended LORAKS to the context of parallel imaging. Besides, Chen *et al.* [34] presented a locally structured low-rank image reconstruction method by imposing low-rank constraints on submatrices of the Hankel structured k-space data matrix.

### C. Score-based Generative Model with SDE

Many deep generative models [35]-[37] have emerged for accelerating MRI reconstruction. Among them, score-based generative models [21], [39]-[41] have gained a lot of success in generating realistic and diverse data. Later, the advanced scored-based generative model with stochastic diffusion equations (SDE) [42] defines a forward diffusion process transforming data into noise and generates data from noise by reversing the forward process.

More specifically, one can consider a diffusion process $\{x(t)\}_{t=0}^T$ with $x(t) \in \mathbb{R}^n$, where $t \in [0,T]$ is a continuous time variable and $n$ denotes the image dimension. By choosing $x(0) \sim p_0$ and $x(T) \sim p_T$, $p_0$ is the data distribution and $p_T$ is the prior distribution, the diffusion process can be modeled as the solution of the following SDE:
$$dx = f(x,t)dt + g(t)dw \quad (3)$$
where $f \in \mathbb{R}^n$ is the drift coefficient, and $g \in \mathbb{R}$ is the diffusion coefficient of $x(t)$. $w \in \mathbb{R}^n$ induces Brownian motion.

According to the reversibility of SDE, the reverse process of Eq. (3) can be expressed as another stochastic process:

$$dx = [f(x,t) - g(t)^2 \nabla_x \log p_t(x)]dt + g(t)d\bar{w} \quad (4)$$

where $dt$ is the infinitesimal negative time step, and $\bar{w}$ is a standard Wiener process for time flowing in reverse. The score term $\nabla_x \log p_t(x)$ can be approximated by a learned time-dependent score model $s_\theta(x(t),t)$. The SDE is then solved by some solver procedures, providing the basis for score-based generative modeling with SDE.

## III. METHOD

### A. Characteristics in Hankel Matrix

The cornerstone of the proposed HKGM method is the Hankel matrix, which exploits and displays correlations in multi-coil MRI k-space data. The Hankel matrix constructs multi-coil data into a single data matrix. As shown in Fig. 2, a single data block in the k-space is vectorized into a column in the data matrix [6]. When reversely forming a k-space dataset from a data matrix, multiple anti-diagonal entries are averaged and stored at appropriate k-space locations.

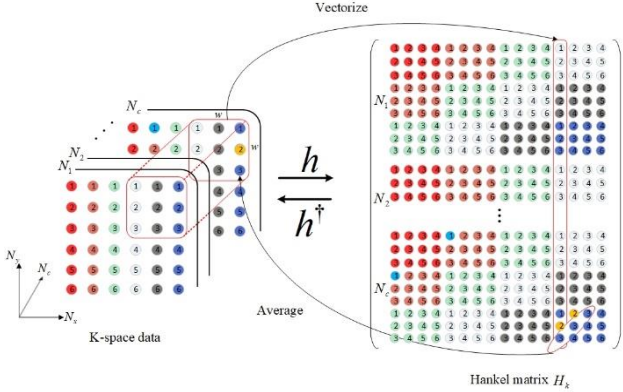

**Fig. 2.** Schematic diagram of the construction of the Hankel matrix.

We define a linear operator $h$ that generates a data matrix $H_k$ from multi-coil k-space data.

$$h: \mathbb{C}^{N_x \times N_y \times N_c} \rightarrow \mathbb{C}^{w^2 N_c \times (N_x - w + 1)(N_y - w + 1)} \quad (5)$$

Then, the reverse operator $h^\dagger$ generates the corresponding k-space dataset from the data matrix $H_k$.

$$h^\dagger: \mathbb{C}^{w^2 N_c \times (N_x - w + 1)(N_y - w + 1)} \rightarrow \mathbb{C}^{N_x \times N_y \times N_c} \quad (6)$$

where $\dagger$ denotes a pseudo-inverse operator. The formulation of Hankel matrix involves two key characteristics:

1) **Redundancy property.** As seen in Fig. 2, any elements in the original matrix can be repeatedly observed in multiple regions of the Hankel matrix. From $N_x \times N_y$ sized data with $N_c$ number of coils, we can generate a data matrix having the size of $w^2 N_c \times (N_x - w + 1)(N_y - w + 1)$ by sliding a $w \times w \times N_c$ window across the entire k-space, so the capacity of Hankel matrix is approximately $w^2 N_c \times (N_x - w + 1)(N_y - w + 1) / N_x \times N_y$ times than that of the original matrix. Therefore, the Hankel matrix is a redundant representation of the original matrix.

2) **Structural property.** Fig. 2 shows a pictorial description of constructing a Hankel matrix with an exemplary $3 \times 3$ window. It should be noted that the data matrix has a stacked, block-wise Hankel structure because of the nature of the sliding-window operation. Due to the linear dependency residing in multi-coil data, this matrix is also low-rank [43]-[45].

In this study, we explore the redundancy and structural properties of the Hankel matrix, similarly as exploring the characteristics of image patches in the past decades [28]. On the one hand, the redundancy property provides the feasibility to extract multiple data patches for prior learning, which can generate relatively sufficient data samples. On the other hand, the structural property guarantees the low-rank feature. It is desirable to utilize the generative model to learn prior information on the patches extracted from the Hankel matrix. Therefore, the internal distribution underlying the single k-space data can be captured for reconstruction.

### B. Prior Learning in Hankel Structured k-Space

Generally speaking, supervised learning approaches require thousands of data samples to train a network. Although generative models aim to learn the probability distribution of the images to be reconstructed by network training, they also require hundreds of data samples. It would be better to get high-quality reconstruction results with only one data sample used for training.

Since direct training with one data sample is intractable, we utilize the Hankel matrix to generate sufficient data samples in this study. Hankel matrix has the feature that can be generated algebraically. A single data block in the k-space is vectorized into a column in the data matrix. Therefore, it can construct a data matrix whose size is much larger than the original matrix. As shown in Fig. 3, we take the fully-sampled k-space data with the size of $256 \times 256 \times 8$ as the input of HKGM to learn prior information. The training process consists of three steps:

**Step 1:** *Constructing a large Hankel matrix.* By sliding a $8 \times 8 \times 8$ window across the entire k-space, we can construct a large block Hankel matrix with the size of $62001 \times 512$, which is much larger than the original matrix. This process can be formulated as:

$$H_k = h(k) \quad (7)$$

**Step 2:** *Extracting k-space patches from Hankel matrix.* Inspired by the success of exploiting the internal statistics of image patches [22]-[30], we randomly extract multiple Hankel structured k-space patches $\{R_k^i\}_{i=1}^N$ with the size of $256 \times 256$ from the Hankel matrix to generate a moderate amount of data samples (e.g., 484 patches). The main purpose of this step is to generate sufficient data samples with similar rank values as the Hankel matrix. As observed from one representative patch in Fig. 4, most singular value coefficients of the patch are near to $0$, which indicates that the patches from the Hankel matrix are indeed low rank. More descriptions regard to the rank between the Hankel matrix and the extracted k-space patches are included in **Supplementary Materials**.

**Step 3:** *Capturing the internal distribution of Hankel structured k-space patches.* After Step 2, we obtain a lot of Hankel structured k-space patches with the same number of elements for training. By constructing the Hankel matrix and extracting patches from it, the same pixel of k-space data exists in different patches. We can fully use the redundancy of the same pixel among different patches to capture the internal statistics underlying the single training k-space data. The internal distribution of each patch is learned through a score-based network. As shown in Eq. (3), the diffusion process in this work can be reformulated as:

$$dR_k = f(R_k,t)dt + g(R_k)dw \tag{8}$$

According to the work of Song *et al.* [42], we use the Variance Exploding (VE) SDE by choosing $f=0$, $g=\sqrt{d[\sigma^2(t)]/dt}$ to form the following Markov chain:

$$R_k^i = R_k^{i-1} + \sqrt{\sigma_i^2 - \sigma_{i-1}^2} z_{i-1}, \ i=1,L,N \tag{9}$$

where $\sigma(t)$ is Gaussian noise function with a variable in continuous time $t \in [0,1]$, which can be redescribed as a positive noise scale $\{\sigma_i\}_{i=1}^N$. Then, we use denoising score matching to minimize Eq. (10):

$$\theta^* = \arg\min_\theta \mathbb{E}_t \{\alpha(t) \mathbb{E}_{R_k(0)} \mathbb{E}_{R_k(t)|R_k(0)}[ \\ \|s_\theta(R_k(t),t) - \nabla_{R_k(t)} \log p_{0t}(R_k(t)|R_k(0))\|^2] \} \tag{10}$$

Here $\alpha:[0,T] \to \mathbb{R}_{>0}$ is a positive weighting function, $t$ is uniformly sampled over $[0,T]$.

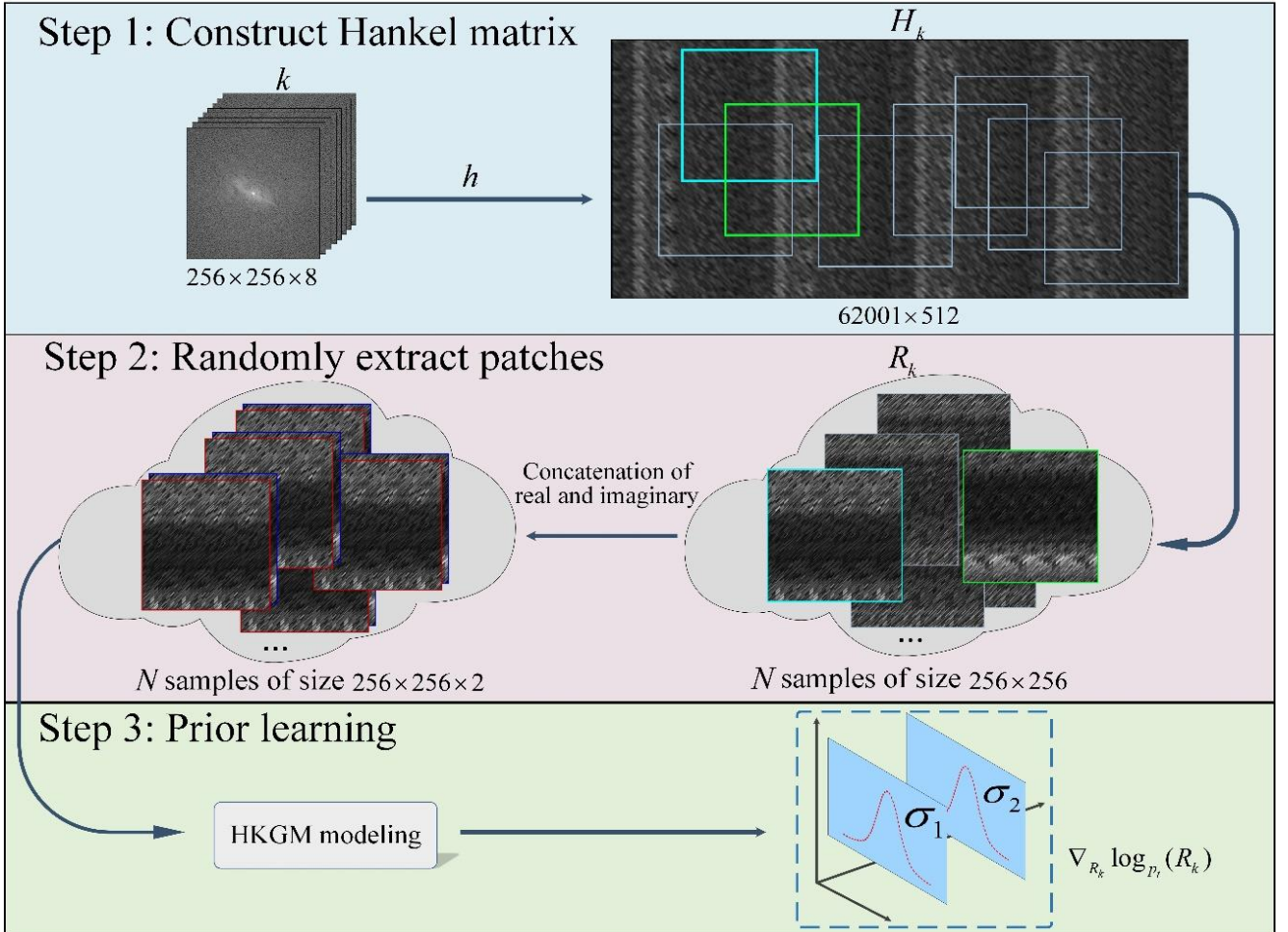

**Fig. 3.** The training flow chart of HKGM. The training process mainly consists of three steps. Firstly, we construct a large Hankel matrix from k-space data. After that, we extract a lot of redundancy and low-rank patches to generate sufficient data samples. Finally, we feed these training patches to the network to capture the internal distribution at different patches.

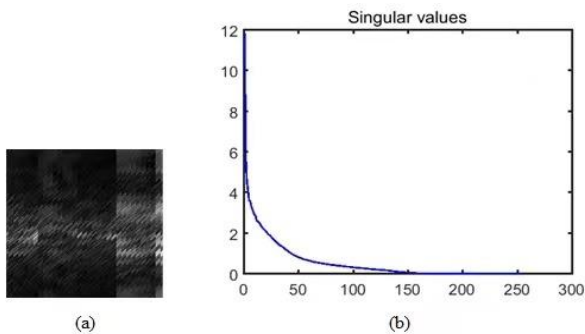

**Fig. 4.** Numerically verification of the low rank property in Hankel structured k-space patches. (a) An example of the Hankel structured k-space patch. (b) Singular value coefficients of the Hankel structured k-space patch.

### C. PMRI Reconstruction of HKGM

By reversing SDE, we can convert random noise into data for sampling and generating high-resolution data. The predictor-corrector (PC) sampling is used for the samples update step. The predictor is a reverse diffusion SDE solver [42], which discretizes the reverse-time SDE in the same way as the forward one. Specifically, the samples from the prior distribution can be obtained from the reverse SDE, which can be discretized as follows:

$$\begin{aligned} k^i &= k^{i+1} + (\sigma_{i+1}^2 - \sigma_i^2)\nabla_k \log p_t(k^{i+1}) + \sqrt{\sigma_{i+1}^2 - \sigma_i^2} z \\ &\approx k^{i+1} + (\sigma_{i+1}^2 - \sigma_i^2) s_\theta(k^{i+1}, \sigma_{i+1}) + \sqrt{\sigma_{i+1}^2 - \sigma_i^2} z \end{aligned} \tag{11}$$

where $i$ is the total number of iterations, $z \sim N(0,1)$ is a zero-mean Gaussian white noise with variance, $k(0) \sim p_0$, and we set $\sigma_0 = 0$ to simplify the notation.

In pMRI reconstruction for under-sampling measurement $y$, Eq. (11) comes to be conditional generation. More precisely, the iterative formulation is as follows:

$$k^i = k^{i+1} + (\sigma_{i+1}^2 - \sigma_i^2)s_\theta(k^{i+1}, \sigma_{i+1}) + (\sigma_{i+1}^2 - \sigma_i^2) \\ [\nabla_k \log p_t(k_{LR}^{i+1}) + \nabla_k \log p_t(y|k^{i+1})] + \sqrt{\sigma_{i+1}^2 - \sigma_i^2}\, z \quad (12)$$

where $\log(p_t(k))$ is given by the prior model that represents information known beforehand about the true model parameter. $\log(p_t(k_{LR}))$ is derived from low-rank data knowledge. $\log p_t(y|k)$ is derived from data knowledge. The detailed derivation of Eq. (12) is provided in **Supplementary Materials**.

The pursuit of Eq. (12) is divided into three subproblems that can be conducted alternatively, i.e., predictor-corrector (PC) operation on the intermediate solution, low-rank penalty on Hankel matrix (i.e., $\nabla_k \log p_t(k_{LR}^{i+1})$) and data consistency (DC) constraint on the measurement (i.e., $\nabla_k \log p_t(y|k^{i+1})$). The whole scheme is shown in Fig. 5. In the following parts, we mainly introduce the solution process of these three subproblems in sequential order.

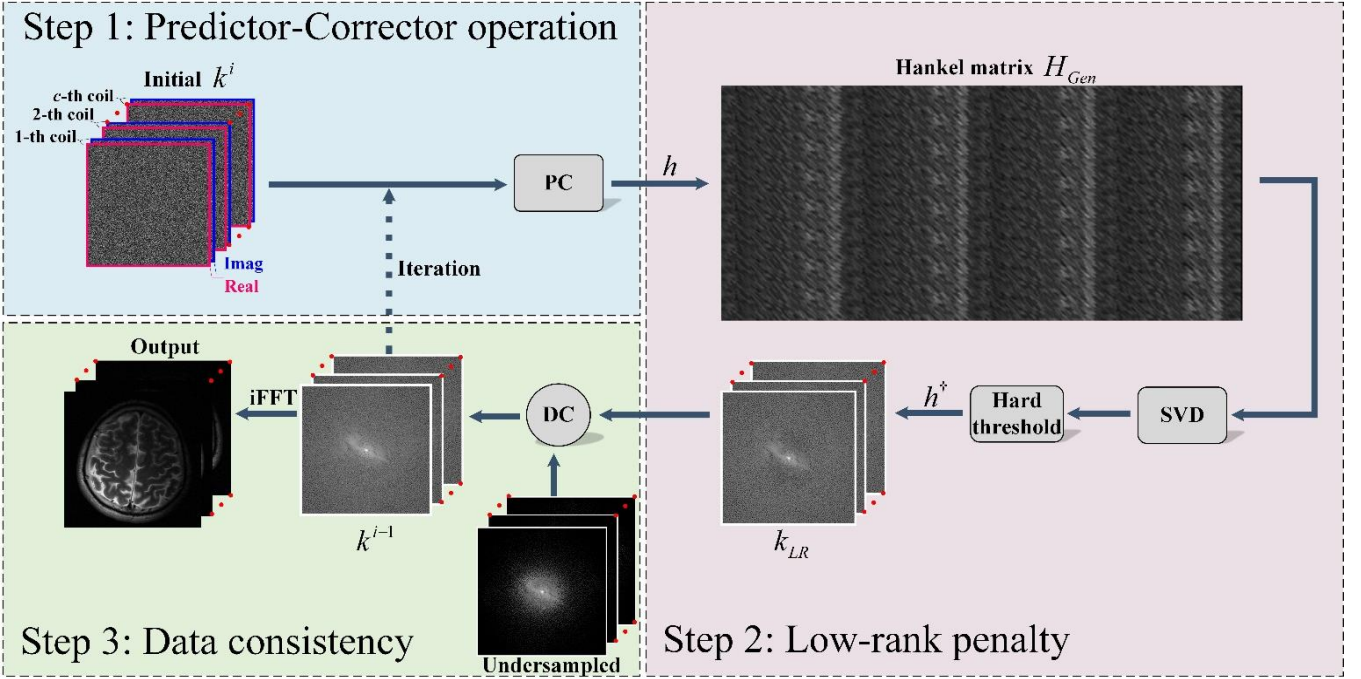

**Fig. 5.** The pipeline of the MRI reconstruction procedure in HKGM. The iterative reconstruction process mainly consists of three steps. Firstly, we iteratively reconstruct objects from the trained network using a PC sampler on the input k-space data. After that, we construct Hankel matrix from the output of the network and apply low-rank penalty on it. Finally, we perform data consistency on the k-space data formed reversely from the matrix.

**Step 1:** *PC.* In the training stage, we extract a lot of Hankel structured k-space patches from the large Hankel matrix and then feed these training patches into the network. During the reconstruction phase, the iterative step would be computationally expensive and time-consuming if we implement it in this manner. To overcome the dilemma, we tend to find another way to make the iterative process more efficient and convenient. It can be observed that k-space data matrix $k$ needs to be interpolated and follows into the prior distribution of Hankel structured k-space patches. As shown in the following formula:

$$k \sim P(R_k) \quad (13)$$

where $P(R_k)$ is the prior distribution of Hankel structured k-space patches. This feature could be utilized to speed up iterations during the reconstruction phase. At each iteration step, we directly use the result of the previous step as the input to the generative model. From Eq (12), we can obtain the generation process:

$$k_{Gen}^i = k_{Gen}^{i+1} + (\sigma_{i+1}^2 - \sigma_i^2)S_\theta(k_{Gen}^{i+1}, \sigma_{i+1}) + \sqrt{\sigma_{i+1}^2 - \sigma_i^2}\, z \quad (14)$$

According to Song *et al.* [42], Eq. (14) can be viewed as the predictor. Corrector is used after the predictor to obtain a more efficient and robust iterative formulation, thus forming PC sampling. The corrector can be viewed as:

$$k_{Gen}^{i,j} = k_{Gen}^{i,j-1} + \varepsilon_i s_\theta(k_{Gen}^{i,j-1}, \sigma_i) + \sqrt{2\varepsilon_i}\, z \quad (15)$$

where the initial solution $k_{Gen}^0$ can be a total uniform noise or other predefined value. $\varepsilon > 0$ is a step size.

After PC operation, we obtain the k-space data $k_{Gen}$ generated by the network, and then we construct a data matrix from it. This process can be formalized as:

$$H_{Gen} = h(k_{Gen}) \quad (16)$$

**Step 2:** *Low-rank constraint on Hankel matrix.* Supposing the data matrix $H_{Gen}$ to be low-rank, we can apply singular value decomposition (SVD) to it to break the information into signal and noise subspaces. An arbitrary $m \times n$ matrix of rank $l$ can be decomposed as:

$$H_{Gen} = \underset{m \times l\ l \times l\ l \times n}{U\ \Sigma\ V^\dagger} \quad (17)$$

To enforce the low-rankness, we make use of the hard thresholding operator $Th_r(\cdot)$, a nonlinear operator that sets the entries smaller than $r$ to zero [47]:

$$Th_r(f) = \begin{cases} f & if \ |f| \geq r \\ 0 & otherwize \end{cases} \quad (18)$$

We perform the threshold operation on the diagonal elements of $\Sigma$, and the hard threshold singular value of the data matrix constructed is obtained by using low-rankness projection, i.e.,

$$Th_r(H_{Gen}) = U(Th_r(diag(\Sigma)))V^\dagger \quad (19)$$

where $U\Sigma V^\dagger$ represents the SVD of $H_{Gen}$. Finally, we project the data matrix onto the k-space data. This operation is done implicitly by applying $h^\dagger$ to the data matrix. The process can be formulated as a constraint optimization, that is:

$$k_{LR} = h^\dagger(Th_r(H_{Gen})) \quad (20)$$

**Step 3: DC.** From Eq. (12), we can obtain the subproblem regarding to data consistency:

$$\underset{k}{Min}\{\|Mk - y\|_2^2 + \lambda \|k - k_{LR}\|^2\} \quad (21)$$

Data consistency is repeatedly enforced in each iterative reconstruction step to ensure that the output is consistent with the original k-space information. The data consistency problem is solved via:

$$k(j) = \begin{cases} k_{LR}(j), & if \ j \notin \Omega \\ [k_{LR}(j) + \lambda y(j)]/(1+\lambda), & if \ j \in \Omega \end{cases} \quad (22)$$

where $\Omega$ denotes an index set of the acquired k-space samples. $k_{LR}(j)$ represents an entry at index $j$ in k-space. In the noiseless setting (i.e., $\lambda \to \infty$), we replace the $j$-th predicted coefficient with the original coefficient if it has been sampled. Then $k(j)$ go back to Step 1 for iterative reconstruction.

The image is reconstructed by applying the inverse Fourier Transform $I = F^{-1}k(j)$. The final reconstruction is obtained by combining the channels through the sum of squares (SOS). At the conditional generation for iterative reconstruction, both the intermediate steps of data consistency and low-rank penalty are non-expansive, thus the whole iterative procedure guarantees convergence. In summary, Step 1-3 in Section III. B and Section III. C constitute the whole process of HKGM. It compactly represents the training and sampling process of HKGM by **Algorithm 1**.

---

**Algorithm 1: HKGM**

**Training stage**
1: **Dataset:** a single k-space data
2: Construct Hankel matrix $H_k = h(k)$
3: Extract patches $\{R_k^i\}_{i=1}^N$ from the matrix
4: **Training:** Eq. (10)
5: **Output:** Trained $S_\theta(R_k, \sigma)$

**Reconstruction stage**
**Setting:** $N, M$
1: $H^N \sim N(0, \sigma_T^2 I)$
2: For $i = N-1$ to 0 do (Outer loop)
3:  $\tilde{k}^i \leftarrow$ Predictor $(k^{i+1}, \sigma_i, \sigma_{i+1})$
4:  $H_{Gen}^i \leftarrow h(\tilde{k}^i)$ (Hankel matrix)
5:  $[U, \Sigma, V] = SVD(H_{Gen}^i)$ (Perform SVD)
6:  $G^{Th_r}(H_{Gem}^i) = U(Th_r(diag(\Sigma)))V^\dagger$
7:  $\tilde{k}_{LR}^i \leftarrow h^\dagger(G^{Th_r}(H_{Gen}^i))$
8:  Update Eq. (22)
9:  For $j = 1$ to $M$ do (Inner loop)
10:   $\tilde{k}^{i,j} \leftarrow$ Corrector$(\tilde{k}^{i,j-1}, \sigma_i, \varepsilon_i)$
11:   $H_k^{i,j} \leftarrow h(\tilde{k}^{i,j})$ (Hankel matrix)
12:   $[U, \Sigma, V] = SVD(H_{Gen}^{i,j})$ (Perform SVD)
13:   $Th_r(H_{Gem}^i) = U(Th_r(diag(\Sigma)))V^\dagger$
14:   $\tilde{k}_{LR}^{i,j} \leftarrow h^\dagger(Th_r(H_{Gem}^i))$
15:   Update Eq. (22)
16:   $k^{i,j} \leftarrow \tilde{k}^{i,j}$
17:  End for
18: End for
19: $k^{rec} \leftarrow \tilde{k}^0$
20: Return $I_{SOS} = \sqrt{\sum_{c=1}^C |(F^{-1}k_c^{rec})|^2}$

---

## IV. EXPERIMENTS

### A. Experiment Setup

**1) Datasets.** First, we used brain images from *SIAT* dataset, which was provided by Shenzhen Institutes of Advanced Technology, the Chinese Academy of Sciences. Informed consents are obtained from the imaging subjects in compliance with the institutional review board policy. The raw data are acquired from 3D turbo spin-echo (TSE) sequence with T2 weighting by a 3.0T whole-body MR system (SIEMENS MAGNETOM Trio Tim), which has 192 slices per slab, and the thickness of each slice is $0.86 \ mm$. Typically, the field of view and voxel size are $220 \times 220 \ mm^2$ and $0.9 \times 0.9 \times 0.9 \ mm^3$, respectively. The relevant imaging parameters encompass the size of image acquisition matrix is $256 \times 256$, echo time (TE) is $149 \ ms$, repetition time (TR) is $2500 \ ms$. Moreover, the number of coils is 12 and the collected dataset includes 500 2D complex-valued MR images. At the training stage, we select one of these images for prior learning, as shown in Fig. 6(a). In addition, we selected 20 of these 500 images as experimental test datasets and named them *Brain1, Brain2,⋯, Brain20*, respectively (excluding the images used for training). We selected five data shown in Fig. 6(d) for reconstruction quality comparison.

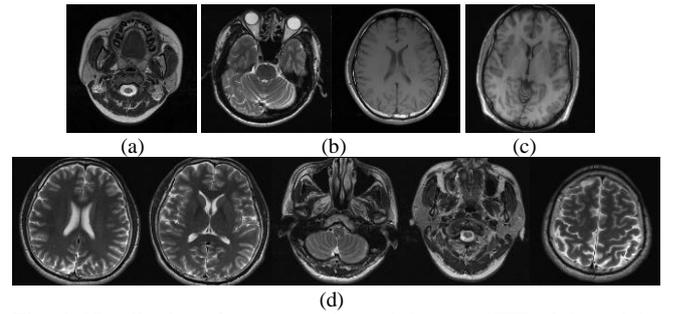

**Fig. 6.** Visualization of some experimental data. (a) iFFT of the training data, (b) 12-channel *T2 transversal Brain* and 8-channel *T1 GE Brain* from [46], (c) 8-channel *T1 Ori Brain* from [6], (d) Five data from SIAT dataset.

Aside from the *SIAT* dataset, we conducted experiments on two in-vivo datasets to verify the reconstruction performance. On one hand, two data depicted in Fig. 6(b) are collected from ref. [46]. One is the 12-channel *T2 transversal Brain* MR images with the size of $256 \times 256$ that acquired with 3.0T Siemens, whose FOV is $200 \times 200 \ mm^2$ and TR/TE is $5000/91 \ ms$ [46]. Another is the 8-channel *T1 GE Brain* with the size of $220 \times 256$ [46]. The MR images are acquired by 3.0T GE. The FOV is $220 \times 220 \ mm^2$, and

TR/TE is 700/11 $ms$. On the other hand, *T1 Ori Brain* in Fig. 6(c) is acquired with a T1-weighted 3D spoiled gradient echo sequence. A single axial slice is used in the experiments [6]. The scan is performed on a 1.5T MRI scanner (GE, Waukesha, Wisconsin, USA) using an 8-channel receive-only head coil. Scan parameters are as: TE = 8 $ms$, TR = 17.6 $ms$, flip angle = $20°$, FOV = $20cm \times 20cm \times 20cm$, and matrix size is $200 \times 200 \times 200$ for an isotropic 1 $mm^3$ resolution.

**2)** *Evaluation Metrics.* The commonly used indexes peak signal to noise ratio (PSNR) and structural similarity (SSIM) are included in the reconstruction experiments to evaluate reconstruction quality. For the convenience of reproducibility, the source code and some representative results are available at: *https://github.com/yqx7150/HKGM.*

**3)** *Model Training and Setting.* The batch size in HKGM is set to be 2 and Adam optimizer with $\beta_1$=0.9 and $\beta_2$=0.999 is utilized to optimize the network. For noise variance schedule, we fix $\sigma_{max} = 1$, $\sigma_{min} = 0.01$ and $r = 0.075$. We set $N = 1000, M = 1$ in all experiments. For the other parameters, we follow the settings in the work of Song *et al* [42]. The training and testing experiments are performed with 2 NVIDIA TITAN GPUs, 12 GB.

### B. Reconstruction Comparisons

We conduct the experiments under various sampling patterns (e.g., Poisson, 2D Partial Fourier and 2D Random sampling) and acceleration factors (i.e., 3×, 4×, 6×, 8× and 10× factors of acceleration). Meanwhile, several state-of-the-arts including ESPIRiT [5], LINDBERG [46], P-LORAKS [35] and SAKE [6] are compared with HKGM.

**1)** *Test on 12-coil T2 SIAT Data.* All methods are evaluated on five data in Fig. 6(d) under different sampling patterns with varying accelerating factors. Fig. 7 presents the reconstructed results at *R*=4 using Poisson sampling with 12 coils. For a closer comparison, we zoom in the image (enclosed with the green box), which implies that the proposed method preserves more details and produces an image closer to the ground truth. The visual quality of reconstructions for different methods varies. ESPIRiT and LINDBERG suffer from artifact, while P-LORAKS, SAKE and the proposed method produce better reconstruction results. There still exist noisy artifacts in P-LORAKS. As for SAKE, the residual map implies that the reconstructed image is deficient in preserving the structure and texture. It can be seen that the proposed HKGM produces the least error.

In addition to the visual comparison, we further provide the average quantitative measurement comparisons for these methods in Table I. The proposed HKGM has the highest PSNR values under both sampling patterns. PSNR of HKGM gains over SAKE from 32.87 dB to 34.76 dB under Poisson sampling pattern with the acceleration factor *R*=4. Similar to the phenomenon in visual comparison, it concludes that HKGM has obtained superior performances quantitatively.

TABLE I
PSNR AND SSIM COMPARISON WITH STATE-OF-THE-ART METHODS UNDER DIFFERENT SAMPLING PATTERNS WITH VARYING ACCELERATE FACTORS.

| Data in Fig. 6(d) | ESPIRiT | LINDBERG | P-LORAKS | SAKE | HKGM |
|---|---|---|---|---|---|
| Poisson R=4 | 31.12/0.833 | 32.20/0.902 | 31.66/0.841 | 32.87/0.866 | **34.76/0.910** |
| Poisson R=10 | 26.68/0.777 | 26.79/0.807 | 28.69/0.738 | 30.13/0.803 | **30.54/0.818** |
| 2D Random R=4 | 30.36/0.775 | 31.18/0.890 | 31.25/0.823 | 32.99/0.829 | **33.87/0.893** |
| 2D Random R=6 | 29.47/0.741 | 28.83/0.850 | 27.95/0.729 | 31.13/0.828 | **31.76/0.858** |

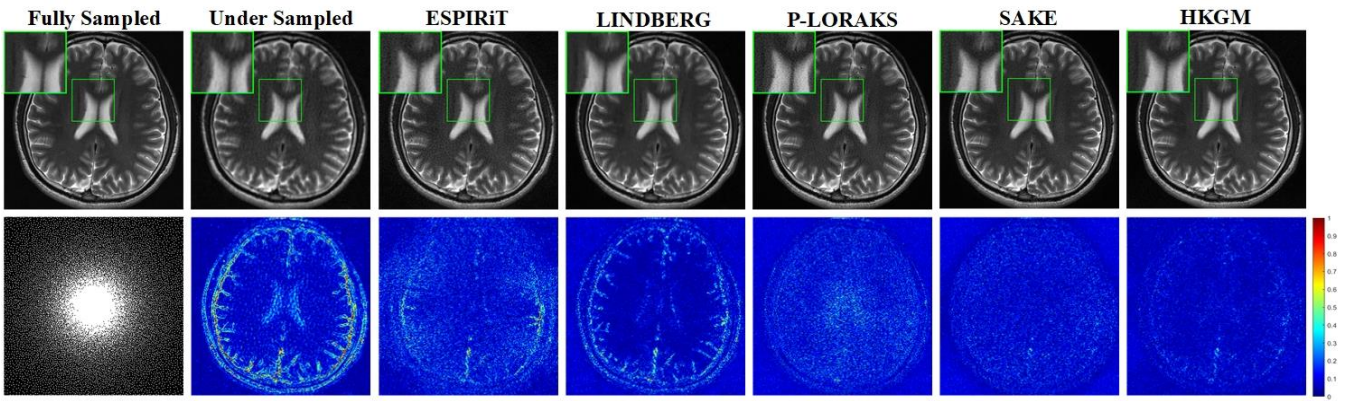

**Fig. 7.** Complex-valued pMRI reconstruction results at *R*=4 using Poisson sampling with 12 coils. From left to right: Fully-sampled, Under-sampled, ESPIRiT, LINDBERG, P-LORAKS, SAKE, and HKGM. The intensity of residual maps is five times magnified.

**2)** *Test on 12-coil T2 Data [46].* Regarding the T2 Transversal Brain data, a Poisson sampling pattern with acceleration factors of 4 and 10 is employed for testing. Fig. 8 visualizes the representative reconstruction results and error maps of HKGM and competing methods. HKGM can produce a realistic reconstruction similar to the ground truth and maintain good characteristics in terms of edge details. The absolute difference maps further illustrate that the competing methods produce more artifacts on the edge or background. ESPIRiT and SAKE exhibit significant noise-like residuals associated with noise vulnerability at high accelerations. LINDBERG exists spatial blurring, which is effectively mitigated by HKGM reconstruction. It can be concluded that HKGM yields superior performance even at an extremely high acceleration factor.

Correspondingly, quantitative evaluations of PSNR and SSIM metrics are listed in Table II. HKGM outperforms ESPIRiT, LINDBERG, P-LORAKS and SAKE in terms of PSNR and SSIM metrics. It is worth noting that the performance of HKGM is optimum under various sampling strategies. HKGM reduces the sampling-specific effects, i.e., the differences between reconstructions with different sampling schemes are less severe. In short, the proposed HKGM effectively improves iteration reconstruction quality and can be generalized to various sampling patterns with different acceleration factors.

TABLE II
PSNR AND SSIM WITH STATE-OF-THE-ART PMRI METHODS UNDER POISSON WITH VARYING ACCELERATE FACTORS.

| *T2 Transversal Brain* | ESPIRIT | LINDBERG | P-LORAKS | SAKE | HKGM |
|---|---|---|---|---|---|
| Poisson *R*=4 | 31.74/0.819 | 32.87/0.901 | 31.44/0.844 | 33.91/0.896 | **35.21/0.921** |
| Poisson *R*=10 | 28.95/0.798 | 26.17/0.822 | 28.96/0.761 | 29.75/0.823 | **31.69/0.850** |

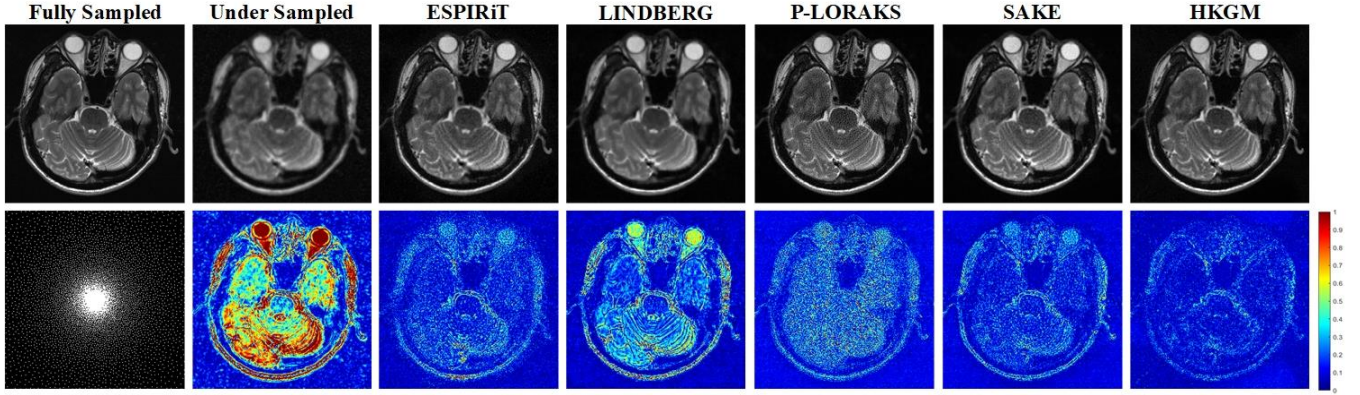

**Fig. 8.** Complex-valued pMRI reconstruction results at *R*=10 using Poisson sampling with 12 coils. From left to right: Fully-sampled, Under-sampled, ESPIRiT, LINDBERG, P-LORAKS, SAKE, and HKGM. The intensity of residual maps is five times magnified.

**3) *Test on 8-coil T1 Data [6].*** As the prior knowledge of HKGM is learned from extracted k-space patches, the learned model from 12-coil object can be applied to reconstruction task with any coils. Table III summarizes the quantitative results of T1-weighted 8-coil data under Poisson sampling pattern with acceleration factors *R*=4, 8 and 2D Partial Fourier sampling pattern with acceleration factors *R*=3, 6, respectively. Bold numbers in Table III indicate the better performance of the proposed method than the competing ones. It can be observed that the average PSNR and SSIM values of the reconstructed each image by using HKGM are higher than those of the other models. Particularly, the performance of HKGM is more striking with the increase in acceleration factor. Even in the case of acceleration factor *R*=8 under Poisson sampling pattern, HKGM still produces reasonable results. Additionally, the SSIM value of HKGM is closer to 1 with Poisson sampling of *R*=4, which indicates that HKGM has good performance in the metric of SSIM between the estimated reconstruction result and ground truth.

Besides the quantitative comparison, the visual quality is also highlighted. Fig. 9 depicts the qualitative reconstruction results of ESPIRiT, LINDBERG, P-LORAKS, SAKE and HKGM, along with their corresponding 5× magnified residual images. From Fig. 9, we can observe that significant residual artifacts and amplified noise exist in the results obtained by ESPIRIT, LINDBERG, and P-LORAKS. SAKE produces an estimation superior to them but still suffers from some visible blurring artifacts. The reconstructed image of HKGM retains more textural details and has the least noise relative to the reference image. To sum up, the proposed HKGM provides realistic reconstruction quality and preserves detailed structures and textures.

TABLE III
PSNR AND SSIM COMPARISON WITH STATE-OF-THE-ART METHODS UNDER DIFFERENT SAMPLING PATTERNS WITH VARYING ACCELERATE FACTORS.

| *T1 Ori Brain* | ESPIRiT | LINDBERG | P-LORAKS | SAKE | HKGM |
|---|---|---|---|---|---|
| Poisson *R*=4 | 36.82/0.886 | 36.97/0.925 | 36.38/0.915 | 38.22/0.933 | **38.92/0.948** |
| Poisson *R*=8 | 34.99/0.837 | 32.13/0.876 | 33.46/0.824 | 35.84/0.881 | **36.46/0.889** |
| 2D Partial *R*=3 | 30.12/0.838 | 30.66/0.919 | 28.97/0.888 | 30.87/0.904 | **31.64/0.927** |
| 2D Partial *R*=6 | 29.68/0.810 | 27.39/0.856 | 28.37/0.826 | 30.18/0.862 | **31.32/0.866** |

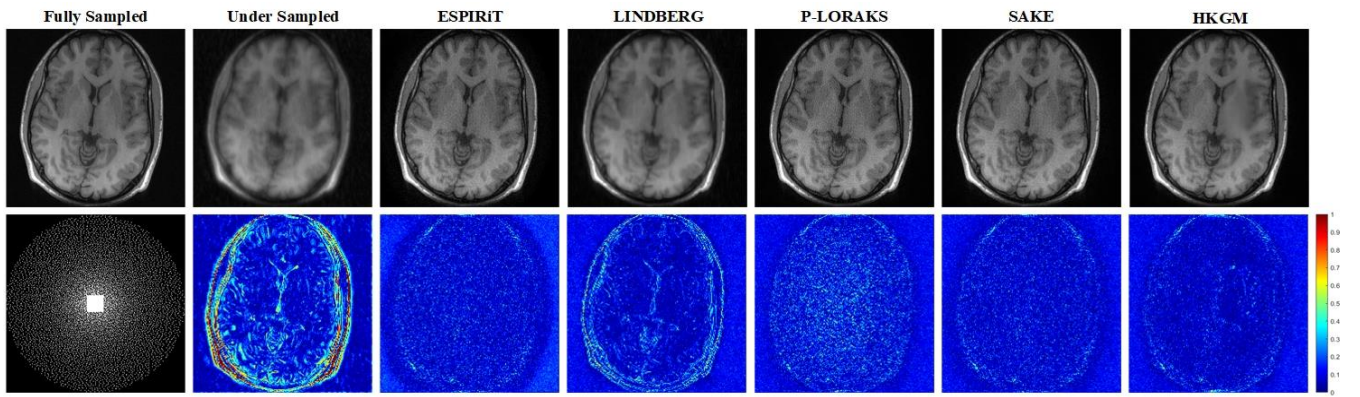

**Fig. 9.** Reconstruction comparison on 2D Poisson sampling at acceleration factor *R*=8. From left to right: Fully-sampled, Under-sampled, ESPIRiT, LINDBERG, P-LORAKS, SAKE, HKGM. The intensity of residual maps is five times magnified.

## C. Convergence Analysis and Computational Costs

In this section, we experimentally investigate the convergence of HKGM and SAKE with the number of iterations. We randomly select an example of reconstructing the T1 GE Brain image in Fig. 6(b) using 2D Random sampling pattern with acceleration factor $R$=6. In Fig. 10, the SSIM curve of HKGM reaches convergence at a fast pace. Simultaneously, the SSIM value of HKGM is always higher than SAKE. In addition, the PSNR curves of both HKGM and SAKE rise rapidly with the increase of iteration. The PSNR curve of SAKE rises more rapidly in the early iterations. However, as the number of iterations increases, the PSNR of HKGM gradually outperforms that of SAKE. After 100 iterations, PSNR value of SAKE starts to decrease, so in the comparative experiments, we set the iteration index of SAKE to 100. More rigorous discussion of the algorithm analysis can be found in **Supplementary Materials.**

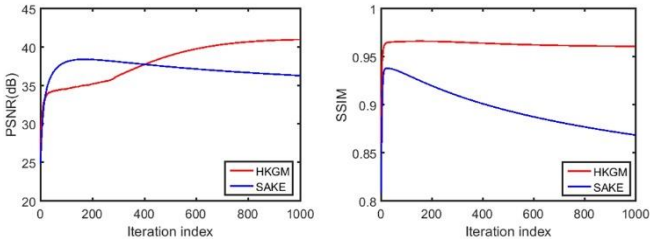

**Fig. 10.** Convergence curves of SAKE and HKGM in terms of PSNR and SSIM versus iterations when reconstructing the brain image from 1/6 2D random sampled data.

## D. Ablation Study

We selected different number of training data samples as the network input to verify the generation capability of HKGM, where two quantitative metrics are listed in Table IV. PSNR and SSIM values attained at $R$=4 Poisson sampling pattern reach the maximum value with 10 images and 100 images as input, respectively. Nevertheless, the difference is very subtle compared to training with only one image. Although the best value for the two quantitative metrics is 100 images as input under the sampling pattern Poisson $R$=6, a reasonable result can be achieved by using only one data for training. In some practical settings, it is difficult to obtain a large number of training data. Therefore, using only one piece of data for training is satisfactory under a comprehensive consideration.

We verify the performance of HKGM by varying the sliding window size at the prior learning stage. In the experiment, the sliding window size varies from $6\times6$, $8\times8$ to $10\times10$. The quantitative results are presented in Table V. As can be seen, both PSNR and SSIM values under the Poisson sampling pattern at $R$=4 and $R$=6 are best at the sliding window size of $8\times8$. In general, the larger the sliding window, the larger the corresponding Hankel matrix. Subsequently, more data patches can be extracted from it. While the sliding window size also affects the rank value of the extracted k-space patches. A larger sliding window results in a higher rank. Therefore, the reconstruction result is best when the sliding window size is $8\times8$.

TABLE IV
PSNR AND SSIM COMPARISON WITH DIFFERENT NUMBER OF INPUT IMAGES UNDER POISSON WITH VARYING ACCELERATE FACTORS.

| Brain1 | 1 | 10 | 100 |
|---|---|---|---|
| Poisson $R$=4 | 36.04/0.920 | **36.09/0.921** | 36.08/**0.921** |
| Poisson $R$=6 | 33.58/0.889 | 33.69/0.891 | **33.97/0.893** |

TABLE V
COMPARISON OF POISSON SAMPLING PATTERNS FOR PSNR AND SSIM UNDER DIFFERENT SLIDING WINDOW SIZES AT THE PRIOR LEARNING STAGE.

| Brain2 | $6\times6$ | $8\times8$ | $10\times10$ |
|---|---|---|---|
| Poisson $R$=4 | 34.67/0.897 | **35.17/0.917** | 35.10/0.915 |
| Poisson $R$=6 | 32.67/0.853 | **32.84/0.879** | 32.79/0.877 |

Table VI quantitatively records the influence of different singular value thresholding on our reconstruction results when the sliding window size is $8\times8$. It can be found that the PSNR value reaches the maximum value when the singular value thresholding is 0.8 under the Poisson sampling pattern. Additionally, both PSNR and SSIM values get the maximum when the singular value thresholding is 0.8 at the Random sampling pattern with acceleration factor $R$=4. The visual quality is included in **Supplementary Materials.**

We fix the singular value thresholding to be 0.8 and verify the performance of the proposed method by varying the sliding window size and sampling patterns at the iterative reconstruction stage. As can be seen from Table VII, as the sliding window size increases, the PSNR becomes higher while the SSIM becomes lower. Under the sampling pattern of Poisson $R$=4, when the sliding window size is $2\times2$, SSIM reaches the maximum value of 0.927. Under the Random $R$=4 sampling pattern, SSIM reaches the maximum value of 0.916 for the window size of $4\times4$. Besides, PSNR reaches the highest value for the window size of $10\times10$ under both Poisson and 2D Random sampling patterns. This phenomenon illustrates that the appropriate window size plays a significant role in the performance of the proposed HKGM.

TABLE VI
PSNR AND SSIM COMPARISON WITH DIFFERENT SAMPLING UNDER VARYING THRESHOLD VALUES.

| Brain3 | $Th_{0.4}$ | $Th_{0.8}$ | $Th_{1.2}$ | $Th_{1.6}$ | $Th_{2.0}$ |
|---|---|---|---|---|---|
| Poisson $R$=4 | 33.84/**0.927** | **34.61**/0.917 | 33.71/0.902 | 34.22/0.907 | 34.20/0.913 |
| 2D Random $R$=4 | 31.20/0.913 | **33.28/0.914** | 32.73/0.892 | 32.80/0.894 | 33.12/0.890 |

TABLE VII
PSNR AND SSIM COMPARISON WITH DIFFERENT SAMPLING PATTERNS UNDER VARYING SLIDING WINDOW SIZES AT THE ITERATIVE RECONSTRUCTION STAGE.

| Brain3 | $2\times2$ | $4\times4$ | $6\times6$ | $8\times8$ | $10\times10$ |
|---|---|---|---|---|---|
| Poisson $R$=4 | 32.66/**0.927** | 33.93/0.925 | 34.11/0.920 | 34.61/0.917 | **34.93**/0.913 |
| 2D Random $R$=4 | 30.49/0.897 | 32.56/**0.916** | 32.40/0.915 | 33.28/0.914 | **33.83**/0.909 |

## V. CONCLUSION

This work introduced a Hankel-k-space generative model that captured the internal statistics of a single training k-space data. At the prior learning stage, we constructed a large Hankel matrix and extracted multiple k-space patches from it to capture the internal distribution underlying the single k-space data. At the iterative reconstruction stage, in addition to iterative generation, we impose the low-rank penalty on the Hankel matrix and data consistency on the measurement data. Experimental results verified that HKGM could produce better performance under different sampling patterns with large acceleration factors, and retained higher PSNR and SSIM values compared to state-of-the-arts.

REFERENCES

[1] M. A. Griswold, M. Blaimer, F. Breuer, et al., "Parallel magnetic resonance imaging using the GRAPPA operator formalism," *Magn. Reson. Med.*, vol. 54, no. 6, pp. 1553-1556, 2005.

[2] M. A. Griswold, P. M. Jakob, R. M. Heidemann, M. Nittka, V. Jellus, J. Wang, B. Kiefer, and A. Haase, "Generalized autocalibrating partially parallel acquisitions (GRAPPA)," *Magn. Reson. Med.*, vol. 47, no. 6, pp. 1202-1210, 2002.

[3] K. P. Pruessmann, M. Weiger, M. B. Scheidegger, and P. Boesiger, "SENSE: Sensitivity encoding for fast MRI," *Magn. Reson. Med.*, vol. 42, no. 5, pp. 952-962, 1999.

[4] M. Lustig, D. L. Donoho, J. M. Santos, et al., "Compressed sensing MRI," *IEEE Signal Process.* Mag., vol. 25, no. 2, pp. 72-82, 2008.

[5] M. Uecker, P. Lai, M. J. Murphy, et al., "ESPIRiT—an eigenvalue approach to autocalibrating parallel MRI: where SENSE meets GRAPPA," *Magn. Reson. Med.*, vol. 71, no. 3, pp. 990-1001, 2014.

[6] P. J. Shin, P. E. Larson, M. A. Ohliger, et al., "Calibrationless parallel imaging reconstruction based on structured low-rank matrix completion," *Magn. Reson. Med.*, vol. 72, no. 4, pp. 959-970, 2014.

[7] J. P. Haldar, "Low-rank modeling of local k-space neighborhoods (LORAKS) for constrained MRI," *IEEE Trans. Med. Imag.*, vol. 33, pp. 668-681, 2014.

[8] K. H. Jin, D. Lee, J. C. Ye, "A general framework for compressed sensing and parallel MRI using annihilating filter based low-rank hankel matrix," *IEEE Trans. Compu. Imag.*, vol. 2, no. 4, pp. 480-495, 2016.

[9] K. Hammernik, T. Klatzer, E. Kobler, et al., "Learning a variational network for reconstruction of accelerated MRI data," *Magn. Reson. Med.*, vol. 79, no. 6, pp. 3055-3071, 2018.

[10] J. Schlemper, J. Caballero, J. V. Hajnal, A. Price, and D. Rueckert, "A deep cascade of convolutional neural networks for MR image reconstruction," in *Proc. Int. Conf. Inf. Process. Med. Imag.*, pp. 647-658, May. 2017.

[11] Y. Liu, Q. Liu, M. Zhang, Q. Yang, S. Wang, and D. Liang, "IFR-Net: Iterative feature refinement network for compressed sensing MRI," *IEEE Trans. Comput. Imag.*, vol. 6, pp. 434-446, 2019.

[12] C. Qin, J. Schlemper, J. Caballero, A. N. Price, J. V. Hajnal, D. Rueckert, "Convolutional recurrent neural networks for dynamic MR image reconstruction," *IEEE Trans. Med. Imag.*, vol. 38, no. 1, pp. 280-290, 2018.

[13] Y. Arefeen, O. Beker, J. Cho, et al. "Scan-specific artifact reduction in k-space (SPARK) neural networks synergize with physics-based reconstruction to accelerate MRI," *Magn. Reson. Med.*, vol. 87, no. 2, pp. 764-780, 2022.

[14] M. Akçakaya, S. Moeller, S. Weingärtner, et al., "Scan-specific robust artificial-neural-networks for k-space interpolation (RAKI) reconstruction: Database-free deep learning for fast imaging," *Magn. Reson. Med.*, vol. 81, no. 1, pp. 439-453, 2019.

[15] T. H. Kim, P. Garg, J. P. Haldar, et al., "LORAKI: Autocalibrated recurrent neural networks for autoregressive MRI reconstruction in k-space," *arXiv preprint arXiv*:1904.09390, 2019.

[16] M. Zhang, M. Li, J. Zhou, et al., "High-dimensional embedding network derived prior for compressive sensing MRI reconstruction," *Medical Image Analysis*, vol. 65, 2020.

[17] Q. Liu, Q. Yang, H. Cheng, S. Wang, M. Zhang, and D. Liang, "Highly undersampled magnetic resonance imaging reconstruction using autoencoding priors," *Magn. Reason. Med.*, vol. 83, no. 1, pp. 322-336, 2020.

[18] K. C. Tezcan, C. F. Baumgartner, R. Luechinger, K. P. Pruessmann, and E. Konukoglu, "MR image reconstruction using deep density priors," *IEEE Trans. Med. Imag.*, vol. 38, pp. 1633-1642, 2018.

[19] G. Luo, N. Zhao, W. Jiang, E.S. Hui, and C. Peng, "MRI reconstruction using deep Bayesian estimation," *Magn. Reason. Med.*, vol. 84, no. 4, pp. 2246-2261, 2020.

[20] V. A. Kelkar, S. Bhadra, M. A. Anastasio, "Compressible latent-space invertible networks for generative model-constrained image reconstruction," *IEEE Trans. Compu. Imag.*, vol. 7, pp. 209-223, 2021.

[21] C. Quan, J. Zhou, Y. Zhu, Y. Chen, S. Wang, D. Liang, and Q. Liu, "Homotopic gradients of generative density priors for MR image reconstruction," *IEEE Trans. Med. Imag.*, vol. 40, no. 12, pp. 3265-3278, 2021.

[22] K. Dabov, A. Foi, V. Katkovnik, et al., "Image denoising by sparse 3-D transform-domain collaborative filtering," *IEEE Trans. Image Process.*, vol. 16, no. 8, pp. 2080-2095, 2007.

[23] W. Zhao, Y. Lv, Q. Liu, et al., "Detail-preserving image denoising via adaptive clustering and progressive PCA thresholding," *IEEE Access.*, vol. 6, pp. 6303-6315, 2017.

[24] M. Elad, M. Aharon, "Image denoising via sparse and redundant representations over learned dictionaries," *IEEE Trans. Image Process.*, vol. 15, no. 12, pp. 3736-3745, 2006.

[25] W. Dong, L. Zhang, G. Shi, X. Li, "Nonlocally centralized sparse representation for image restoration," *IEEE Trans. Image Process.*, vol. 22, no. 4, pp. 1620-1630, 2012.

[26] J. Yang, J. Wright, T. S. Huang, et al., "Image super-resolution via sparse representation," *IEEE Trans. Image Process.*, vol. 19, no. 11, pp. 2861-2873, 2010.

[27] Y. Bahat, M. Irani, "Blind dehazing using internal patch recurrence," *IEEE International Conference on Computational Photography*, pp. 1-9, 2016.

[28] T. S. Cho, M. Butman, S. Avidan, et al., "The patch transform and its applications to image editing," *IEEE Conference on Computer Vision and Pattern Recognition*, pp. 1-8, 2008.

[29] T. H. Vu, H. S. Mousavi, V. Monga, et al., "Histopathological image classification using discriminative feature-oriented dictionary learning," *IEEE Trans. Med. Imag.*, vol. 35, no. 3, pp. 738-751, 2015.

[30] A. Buades, B. Coll, J. M. Morel, "A non-local algorithm for image denoising," *IEEE computer society conference on computer vision and pattern recognition.*, vol. 2, pp. 60-65, 2005.

[31] Y. Han, L. Sunwoo, J. C. Ye, "k-Space deep learning for accelerated MRI," *IEEE Trans. Med. Imag.*, vol. 39, no. 2, pp. 377-386, 2019.

[32] X. Zhang, D. Guo, Y. Huang, et al., "Image reconstruction with low-rankness and self-consistency of k-space data in parallel MRI," *Medical Image Analysis*, 2020.

[33] J. P. Haldar, J. Zhuo, "P-LORAKS: low-rank modeling of local k-space neighborhoods with parallel imaging data," *Magn. Reason. Med.*, vol. 75, no. 4, pp. 1499-1514, 2016.

[34] X. Chen, W. Wu, M. Chiew, "Locally structured low-rank MR image reconstruction using submatrix constraints," *IEEE 19th International Symposium on Biomedical Imaging (ISBI)*, pp. 1-4, 2022.

[35] H. Chung, and J. C. Yea, "Score-based diffusion models for accelerated MRI," *arXiv preprint arXiv:2110.05243*, 2021.

[36] A. Jalal, M. Arvinte, Daras G, E. Price, A. Dimakis, and J. Tamir, "Robust compressed sensing MRI with deep generative priors," *Adv. Neural Inf. Process. Syst.*, vol. 34, 2021.

[37] Y. Xie, and Q. Li, "Measurement-conditioned denoising diffusion probabilistic model for under-sampled medical image reconstruction," *arXiv preprint arXiv:2203.03623*, 2022.

[38] P. Vincent, "A connection between score matching and denoising autoencoders," *Neural Comput.*, vol. 23, no. 7, pp. 1661-1674, 2011.

[39] Y. Song, and S. Ermon, "Generative modeling by estimating gradients of the data distribution," *Proc. Adv. Neural Inf. Process. Syst.*, pp. 11895-11907, 2019.

[40] A. Hyvärinen and P. Dayan, "Estimation of non-normalized statistical models by score matching," *J. Mach. Learn. Res.*, vol. 6, no. 4, 2005.

[41] W. Zhu, B. Guan, S. Wang, et al., "Universal generative modeling for calibration-free parallel MR imaging," *IEEE 19th International Symposium on Biomedical Imaging (ISBI)*, pp. 1-5, 2022.

[42] Y. Song, J. Sohl-Dickstein, D. P. Kingma, A. Kumar, S. Ermon, and B. Poole, "Score-based generative modeling through stochastic differential equations," *arXiv preprint arXiv:2011.13456*, 2020.

[43] J. Zhang, C. Liu, M. E. Moseley, "Parallel reconstruction using null operators," *Magn. Reson. Med.*, vol. 66, no. 5, pp. 1241-1253, 2011.

[44] M. Lustig, M. Elad, J. M. Pauly, "Calibrationless parallel imaging reconstruction by structured low-rank matrix completion," *Proceedings of the 18th Annual Meeting of ISMRM*, pp. 2870, 2010.

[45] M. Lustig, "Post-cartesian calibrationless parallel imaging reconstruction by structured low-rank matrix completion," *Proceedings of the 19th Annual Meeting of ISMRM*, pp. 483, 2011.

[46] S. Wang, S. Tan, Y. Gao, Q. Liu, L. Ying, T. Xiao, Y. Liu, X. Liu, H. Zheng, and D. Liang, "Learning joint-sparse codes for calibration-free parallel MR imaging," *IEEE Trans. Med. Imag.*, vol. 37, no. 1, pp. 251-261, 2017.

[47] S. Ma, "Algorithms for sparse and low-rank optimization: Convergence, complexity and applications," Thesis, June 2011.